\documentstyle[prl,twocolumn,epsfig,aps,floats]{revtex}

\begin{document}

\title{ Neutrino-less Double Electron Capture - a tool to 
search for Majorana neutrinos}

\author{Z.Sujkowski
and S. Wycech\thanks{{\it e-mails:} sujkow@fuw.edu.pl, wycech@fuw.edu.pl}}
\address{So{\l}tan Institute for Nuclear Studies,
Ho\.za 69, PL-00-681, Warsaw, Poland}
 
%\author{ S. Wycech\thanks{{\it Internet address:} wycech@fuw.edu.pl}}
% \address{So{\l}tan Institute for Nuclear Studies,
% Ho\.za 69, PL-00-681, Warsaw, Poland}

\maketitle

\begin{abstract}
The possibility to observe the neutrino-less double 
$ \beta $ decay  and thus to prove the  Majorana nature of 
neutrino as well as to provide a sensitive measure of its mass 
is a major challenge of  to-day's neutrino physics. 
As an attractive alternative we propose to study the inverse  
process, the radiative  neutrino-less double 
electron capture $0 \nu 2EC$. The associated monoenergetic  
photon provides a convenient experimental signature. 
Other advantages include the favourable ratio of the  $0 \nu 2EC$ 
to the competing  $2\nu 2EC$ capture rates and, 
very importantly, the existence of coincidence trigger 
to suppress the random background.
These advantages partly offset the expected longer lifetimes.
Rates for the $0\gamma 2EC$ process are calculated. High Z atoms 
are strongly favoured.  A resonance enhancement of the capture 
rates  is expected to occur at energy release comparable to 
the $2P-1S$ atomic level difference. The resonance conditions 
are likely to be met for decays to excited states in final nuclei. 
Candidates for such studies are 4considered.
The experimental feasibility is estimated and found encouraging.\\

\end{abstract}

%\pacs{PACS numbers: 13.15+g,23.40.Bw}

%\section{Introduction}

\sc{Introduction.}
\normalshape
The  existence of  massive neutrinos and the Dirac or Majorana nature
of these particles are among the most intriguing topics  of the present day 
physics. If neutrino is a Majorana particle then  
by definition it  is identical  to its charge conjugate. Thus 
the neutrino  produced in one weak interaction vertex  
may be absorbed in another one.
This leads to the nuclear reaction $0\nu \beta \beta$:
\begin{equation}
\label{1}
 (A,Z-2) \rightarrow (A,Z) +e +e 
\end{equation}  
(see fig.1). Amplitudes for such a process are proportional to the 
Majorana neutrino mass. 
While the exact value of the
neutrino mass deduced from a successful $0\nu \beta \beta$ 
experiment is model dependent, the mere observation of the 
effect proves unambiguously the  Majorana nature of neutrino  
as well as the nonconservation of the lepton number. This 
remains true regardless of the
mechanism causing the decay \cite{MOH01},\cite{BOE87}.

The $0\nu \beta \beta $ process proceeds via the emission 
of two correlated $ \beta $ electrons. 
Its  unique signature  is that  the sum of 
energies of the two  electrons is equal to the total 
decay energy. The experiments searching for the 
${0\nu}{\beta^-}{\beta^-}$ decay can
be divided into two categories: the calorimetric experiments, 
in which the material of the source is usually identical with 
that of the detector, and the tracking experiments, in which 
the source and the detector are separate.  The former 
automatically sums up the energies of the charged particles emitted.
Large quantities of the material can be used. The main difficulty 
rests in suppressing the random background. The only way to attain 
this is by requiring extreme shielding conditions as well as  
the extreme purity of all the material of the detector housing and 
of the surrounding. The tracking detectors, counting the two 
electrons in coincidence, are somewhat less sensitive to the 
background. On the other hand, there are practical difficulties 
in handling the neccessarily large amount of the source  material, 
of the order of tons, in the form of thin sheets sandwiched between 
the detectors.  The detectors in both kinds of experiment must 
fulfill the high resolution requirement. Otherwise
the ${0\nu}{\beta^-}{\beta^-}$ peak in the sum spectrum will 
not be discernible from the dominating continuous physical 
background due to the 2${\nu}{\beta^-}{\beta^-}$ decay.
For the description of the problems involved we refer to 
reviews  \cite{BOE87} - \cite{SUJ03}.

We suggest to study the inverse of neutrino-less double 
$\beta^-$  decay, i.e. the 
neutrino-less double electron capture. 
The excess energy is carried away by a photon.
Crude estimates  \cite{SUJ02} for such a radiative process are  
encouraging,  suggesting feasible experiments. These are discussed 
in the next section  on the basis of a theory of the radiative 
capture adapted to the double electron capture case. The point 
of special interest is  the resonant effect which happens when 
the photon energy equals the energy of atomic $2P-1S$ transition. 
In fortunate situations this  effect may strongly enhance 
the $0 \nu c c$ rates.

There are several   experimental advantages  of the radiative 
electron capture
process :\\
$\bullet$ the monoenergetic photon escapes easily from fairly thick layers of the source material
without energy degradation; \\
$\bullet$ the source can be separate from the detector;\\
$\bullet$ the physical background due to the competing $ 2\nu ee\gamma$ 
process is low \cite{WYC03}; \\
$\bullet$ the photon emission is followed by that of the K X-ray; \\
this provides a precious  coincidence trigger 
to suppress the overwhelming random background;\\
$\bullet$ low decay energies are favoured. Decays to 
excited states can thus be considered in realistic experiments in 
contrast to the $\beta\beta$ decays. The $\gamma$ transitions which 
follow offer yet another characteristic coincidence trigger. 

The price to pay is a sizeable reduction of the transition rates.
The experimental questions and the optimal choice 
of the isotopes are discussed in the final  sections.

 \begin{figure}
  \includegraphics[height=.18\textheight, width=.2\textwidth]{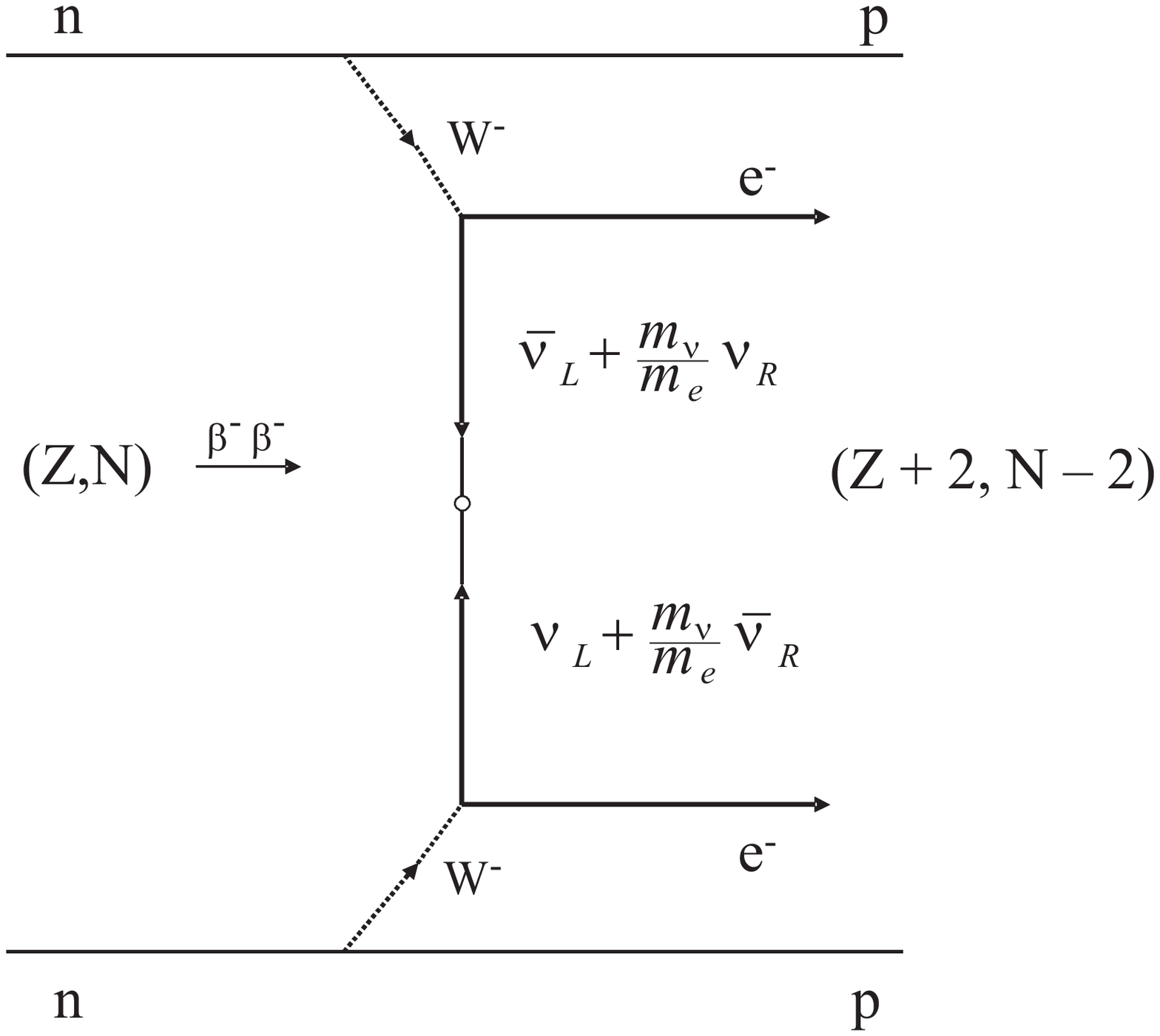}
  \hfill
  \includegraphics[height=.2\textheight, width=.2\textwidth]{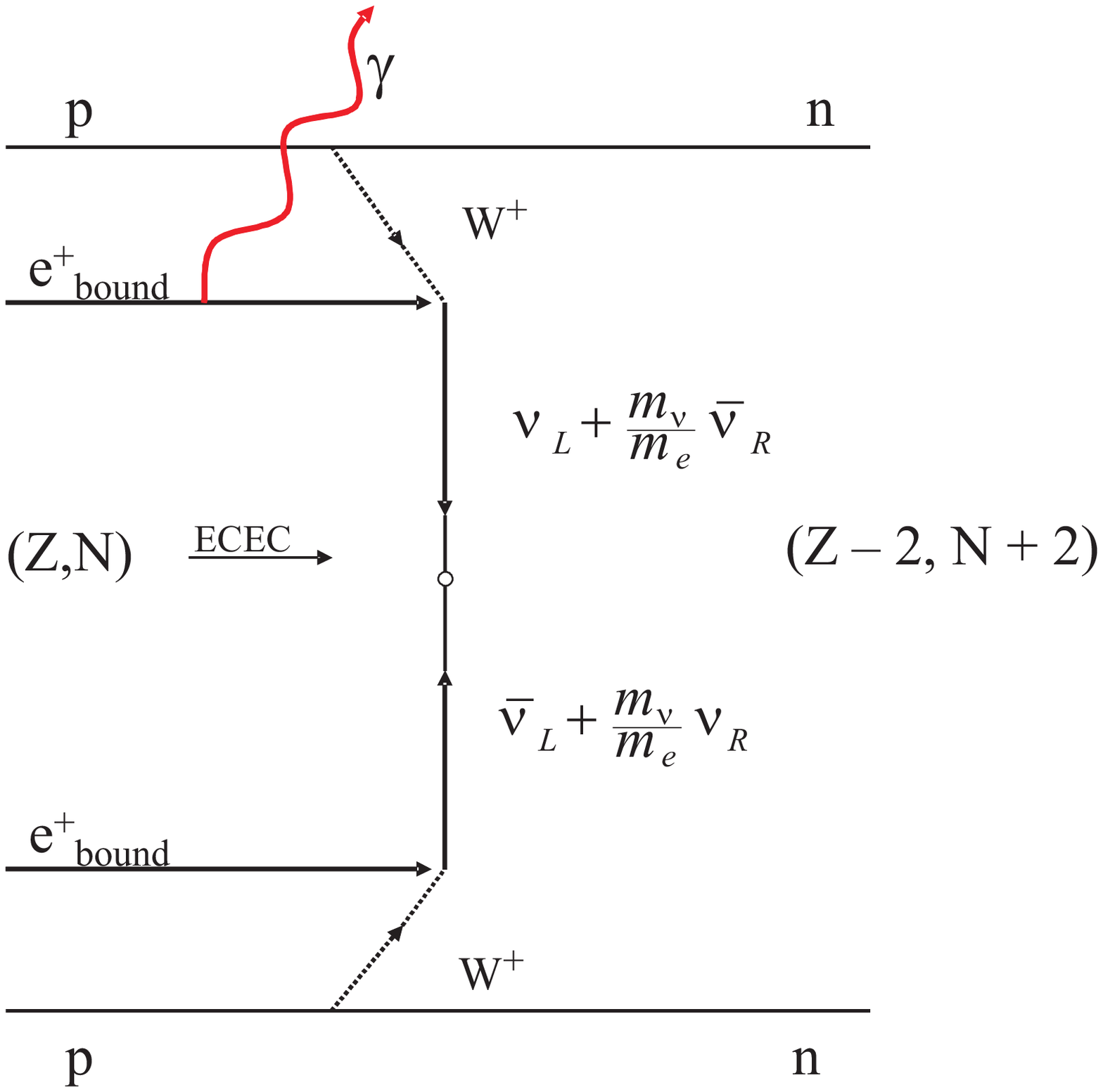}
%  \reflectbox{\includegraphics[height=.3\textheight,angle=-90]{neutron1.eps}}
% \includegraphics[angle=90, height=.2\textheight , width=.2\textwidth]{neutron1.eps}
 \vspace{0.5cm}
 \caption{Diagrams for the $ 0 \nu$ double beta decay and 
 double electron capture processes.} 
 \end{figure}

\sc{Amplitudes for the radiative capture.}
\normalshape
The dominant amplitudes for 
the radiative double electron capture can be factorised into weak-nuclear,
weak-leptonic and radiative factors \cite{DKT85}, \cite{SUH98}.
The expressions for radiative factors 
presented below are based on the theory for the radiative 
single electron capture 
process  \cite{GLA56}, \cite{MAR58}. 

The amplitude for 
$0\nu \beta \beta$  decays  from the $(atom + nucleus)$ ground state 
$\mid 0 )$  to a final nuclear state $N_f$  and final 
two  electron state  $n_f$ is  
\begin{equation}
\label{a1}
 R_{0 \nu}^{ \beta \beta }= 
( 0 \mid  H_w \mid n_f, N_f).
\end{equation}  
This  differs from the amplitude for the 
double electron capture 
\begin{equation}
\label{a2}
 R_{0 \nu}^{ cc }= 
(\bar{n}_f , N_f  \mid H_w \mid  0)
\end{equation}  
by the electron wave functions. Here  
$\bar{n}_f$ denotes two electron holes in the final 
atom  (say $1S,2S$ or $1S,2P$).  
The  reversed transition  cannnot be tested in the same nucleus
and  $ R_{0 \nu}^{ cc }$ does not correspond to 
a real physical transition. 
Assuming the radiative process instead, we can write the corresponding amplitudes
in the second order perturbation theory, to the leading order in the radiative
interaction $H_{\gamma}$, as
\begin{equation}
\label{a3}
R_{0 \nu}^{\gamma}= \Sigma_{n_i, N_i}  
( n_f, N_f \mid H_{\gamma} \mid n_i, N_i)
\frac{( n_i, N_i \mid H_w \mid 0)}{ E - E(n_i) - E(N_i)}.
\end{equation}  
where $H_{w}$ is the weak interaction describing 
the neutrino-less process of fig.1.
Intermediate states $ \mid n_i,N_i)$ in the radiative process may 
involve  ground and excited nuclear states. Only the electron is considered to
radiate as the $\gamma$ emission by the nucleus is unlikely \cite{VER83},
\cite{DKT85}. 

We look first for the transitions which conserve the 
nuclear angular momentum ($ 0^+ \rightarrow 0^+$)  as these 
offer the largest nuclear matrix elements. 
The  final photon and its spin is generated by  the electron  
radiation. Two basic processes are possible:

(a) One of the electrons captured from the  initial atomic state $n_{ini}$ 
radiates while it propagates towards the nucleus.

(b) Two electrons are captured in a virtual process which generates
a final atom with two electron holes. This final atom radiates and
one of the holes is filled. 

Process (a) is standard in the single electron radiative 
captures.  
The description of electron propagation involves the 
electron Coulomb  Green's functions \cite{GLA56},\cite{MAR58}.  
The two electron capture requires some modifications. 
To conserve the angular momentum, for one electron  in  1S state  
the other one must 
be in a higher nS  state. The transition is of
magnetic type. The electron wave functions 
must be  antisymmetrized.     

Process (b) is less important  in general. However, for small 
photon energy $Q$ 
the virtual two electron hole  state  may be degenerate to the final 
atom-photon  state (see below, eq. (\ref{a3a}); see also 
refs.\cite{WIN55}, \cite{GEO81}, \cite{BER83}).  This leads 
to a  singularity in $R_{0 \nu}^{\gamma}$.
The most interesting situation occurs for $Q = E(2P)- E(1S)$ 
i.e. when the final photon resonates with the $ 2P-1S $ transitions 
in the final atom. The capture rate is enhanced up to the limit 
given by the  natural K-L line widths. 
%Possibility  of that type 
%indicated by Winter   as a " monumental coincidence" 
%rediscovered in ref.  were shown 
%in ref. as quite likely to happen. 

The cases (a) and (b) require different  descriptions.
In the simplest theory of Majorana neutrino 
\cite{DKT85},\cite{SUH98} the matrix element $ R_{0 \nu}^{cc}$ is:
\begin{eqnarray}
\label{a5}
R_{0 \nu}^{cc}= 
2\left(\frac {G}{\sqrt{2}} \right)^2 
\int d{\bf x}d{\bf y} 
\langle J_N({\bf x})^{\rho}m_\nu h_{\nu}({\bf x}-{\bf y})
J_N({\bf y})^{\sigma}\rangle \nonumber \\  
\times \Psi(n_1,{\bf x})\gamma_\rho (1-\gamma_5)
\gamma_\sigma  \bar{\Psi}^C(n_2,{\bf y}).
\end{eqnarray}  
The  nuclear weak currents are denoted 
by $J_N$ and the effect of neutrino propagation 
is included into the "neutrino potential" with 
$ h_\nu(r)\approx exp(-qr)/r$, where $q$ is an average momentum carried by 
the neutrino (the closure approximation over nuclear states is used here).
The projection on left-handed intermediate neutrino  
 brings about the neutrino mass factor $m_\nu$. The electronic 
part of this formula contains  atomic wave functions  $\Psi$
and the charge conjugates $\Psi^C$. 
For the $ 0^+ \rightarrow 0^+$
transitions the nuclear part is reduced to Fermi and Gamow-Teller 
matrix elements  defined by 
$ M_F= \langle 0 \mid h_\nu \mid 0 \rangle$ and 
$ M_{GT}= \langle 0\mid h_\nu {\boldmath \sigma}_1 {\boldmath \sigma}_2\mid 0 \rangle$.
These enter via a combination 
$M^{0\nu} =M_{GT}- (\frac {g_V}{g_A})^2 M_{F}$ 
and in this way   the transition matrix element 
(\ref{a2}) is brought to the form 
\begin{eqnarray}
\label{a6}
R_{0 \nu}^{cc}= 
2\left(\frac {G}{\sqrt{2}}\right)^2 M^{0\nu}
 m_\nu  [\Psi(n_1,0)   S(1,2) \Psi(n_2,0)]_A,
\end{eqnarray}  
which resembles the  standard $\beta\beta$ decay expression, \cite{DKT85}.
The electron wave functions are required in the nuclear region. 
These are $\Psi(n_e,0){\hat u}$,  where  
$\Psi(n_e,0)$ is the radial part of the large component and 
${\hat u}$ is the Dirac spinor. The spin matrix element  is given by  
$S(1,2)= u(n_1)(1+\gamma_5)\bar{u}^C(n_2)$. In eq.(\ref{a6}) 
the electron wave function is to be antisymmetrised. For two 
1S electrons this involves antisymmetric spin zero combination of 
spinors which compensates an "inner antisymmetry"  built into operator
$S(1,2)$ by the charge conjugation.  

To describe the  radiative capture  of the type (a), one function 
$ \Psi $ is to be replaced by some  function $\Psi_{\gamma}$ 
which takes care of the photon emission and electron propagation.  
In  eq.(\ref{a3}), in coordinate representation, this function 
is given by the expression  
\begin{equation}
\label{a4}
\Psi_{\gamma}({\bf  r}, n_{ini}) =  \Sigma_{n_i} 
\frac{({\bf  r}\mid  n_i ) ( n_i \mid H_{\gamma} \mid n_{ini})
}{ E - E(n_i) - E(0)}
\end{equation}  
which involves  the sum over continuum and discrete states of 
the electron,  
i.e. the Dirac  Green's function in the external Coulomb 
field of the nucleus (Glauber and Martin  
\cite{GLA56}, \cite{MAR58}). This solution is now implemented into
eq.(\ref{a3}) to give the radiative  amplitude 
\begin{equation}
\label{a6a}
R_{0 \nu}^{\gamma}= 
2\left(\frac {G}{\sqrt{2}}\right)^2 
 m_{\nu} M^{0\nu} \left[\Psi (n_1,0) 
S(1,2, \vec{\epsilon},\vec{Q})_{\gamma} \Psi(n_2,0)_{\gamma}\right]_A
\end{equation}  
where  the leptonic spins and the photon polarisation 
$\epsilon$  enter the last term 
\begin{eqnarray}
\label{a7a}
&&S(1,2, \vec{\epsilon},\vec{Q})_\gamma = \frac{\sqrt{\alpha}}{2m_e} \nonumber \\
&&\times ( u(n_1) [ i A(Q) \vec{\sigma}\vec{\epsilon}\times \vec{Q} -  
B(Q) \vec{\epsilon}\vec{Q} ] (1+\gamma_5)\bar{u}^C(n_2)).
\end{eqnarray}  

For the discussion of $A(Q)$ and $B(Q)$ we refer to ref. \cite{MAR58}. 
If $ Q \approx m_e $ then  $ A , B  $  are close to unity  
for both the 1S and 2S  electrons. The spin conservation 
requires the two electrons to be in the symmetric spin and antysymmetric 
space combinations. This causes cancelations at small $Z$.
% The radiative amplitude  is damped by $ Z\alpha $. 
Otherwise the effect of symmetrisation  is small.    

The rate for radiative no-neutrino process is now 
\begin{equation}
\label{a8}
 \Gamma^{0 \nu \gamma }(Q) = \Sigma_{pol} \int 
\frac{2 \pi d{\bf k }}{(2\pi)^3 2k }\delta(k- Q) 
 \mid R_{0\nu}^\gamma \mid^2,
\end{equation} 
to be summed over photon polarisation and possible 
electron pairs. The result is 
\begin{eqnarray}
\label{a10}
 \Gamma^{0 \nu \gamma} = 
\left(\frac {G}{\sqrt{2}}\right)^4  (M_{GT}- (\frac {g_V}{g_A})^2 M_{F})^2 
(m_\nu/m_e)^2 \nonumber \\  
\times \mid \Psi(1S,0)\Psi(2S,0)\mid ^2 
\frac{Q}{2\pi } \Sigma_{pol} \langle S(n_1,n_2, \vec{\epsilon},\vec{Q})_\gamma\rangle^2 
\end{eqnarray}  
where the summation over spin factor $\langle S\rangle$ is to be taken.
This factor takes care of the angular momentum conservation. 
The best  possibility is  
the $1S,2S$ pair capture accompanied by  a magnetic photon transition. 
 This  rate 
is  scaled by the  $S$ electron  wave function at the nucleus. 
One has 
$ \Psi(n_{1S},0)^2 =(Zm_e\alpha)^3 f^2_{s}/\pi $ where factor 
$f_{s}$ comes from  the weak singularity of 
 Dirac atomic wave function. Averaged over the nuclear volume 
it becomes $ f_{s}= (2R Zm_e \alpha)^{\lambda -1}$ with  
$\lambda= \sqrt{1-(Z\alpha)^2}$ and the nuclear radius $R$. 
The $Z^6$ factor arises and thus one  is interested in the 
heaviest possible atoms. There the 
relativistic singularity enhances  the rate further. 

At smaller $Q$ the capture amplitude  indicates  an interesting structure. 
The  electron propagator in eq.(\ref{a4}) 
has a pole at $Q = \mid E(1S)-E(2P)\mid \equiv Q_{res}$. 
To elucidate the  atomic physics involved in this  radiation process,  
amplitude (\ref{a3}) is presented in terms of the initial  state 
\begin{equation}
\label{a3a}
\mid \Psi_{ini})= \mid 0 )+ \Sigma_{n_i, N_i} 
\frac{ ( n_i, N_i \mid H_w \mid 0)}{ E - E(n_i) - E(N_i)} \mid n_i,N_i) ,
\end{equation}  
where $E = E_N(0)+ E_{atom}(0) $ is the initial energy of 
the system composed of the nuclear part $E_N$ and 
the atomic component $E_{atom}$.
Equation (\ref{a3a}) describes the virtual transitions of the initial 
state $ \mid 0)$ under  weak interactions which include
virtual two electron capture states.  Some of these states 
may be degenerate with the final  $atom + photon$ state 
%What matters in the 
%further  discussion is that the energies involved ammount to 
%some  $100$  keV  and characteristic times  of such virtual 
%transitions are about $ 10 ^{-20} s$.  These are much 
%shorter than the radiation 
%times and shorter than the time required for a rearrangement 
%of the atomic system.  In this sense it seems that the best 
%description of the intermediate states is made in terms 
%of the  initial charge  $Z$.
to within the natural width of the K-L line. Rate enhancements of 
the order of
$10^4 - 10^6$ may be obtained.  
In the single electron description by Glauber and 
Martin the electron propagator in eq.(\ref{a4}) has a pole at 
$Q = \mid E(1S,Z)-E(2P,Z)\mid \equiv Q_{res}(Z)$ where Z refers 
to the initial atom. In the many electron process considered 
here we use the energy levels for ${final (Z - 2)}$ atom
(see also ref. \cite{BER83}).
We obtain 
\begin{equation}
\label{a11}
 \Gamma^{0 \nu \gamma}(Q)= 
\frac {\Gamma^r(2P \rightarrow 1S) }{ [ Q- Q_{res}(Z-2)]^2+ [ \Gamma^r/2 ]^2}
 \mid R_{0\nu}^{cc} \mid^2
\end{equation}
where $\Gamma^r $ is the radiative width of the final 
two electron hole atom.  We use the experimental energy 
values for the $Z-2$ atoms corrected for
the screening differences in the ionised atoms; the combined 
width is $3\Gamma^r(1S)+ \Gamma^r(2P)$ 
similarly as for the hypersatellite $K^h X$-ray spectra.

\sc{Experimental feasibility, optimal targets.}
\normalshape
The usual choice for targets in the double $\beta$ decay 
is motivated by the phase space that favors large energy release  
( roughly $Q_{\beta \beta}^5 $ dependence).
In contrast, the phase space dependence of the double electron 
capture eq.(\ref{a10}) is rather weak 
and the radiative factors  favor small $Q$. 
Two capture situations can be considered.  At  $Q$ larger than 
the electron mass,  the "magnetic" radiative capture of $1S,2S$ 
pair dominates.   
The calculated rates (Table 1)  are based on values 
$ M_{GT} \approx 0.6/fm $ 
and $ M_{F}/ M_{GT} \approx - 0.3$ obtained in typical 
$\beta \beta 0\nu$ reactions in a number of nuclei \cite{SUH01}, 
\cite{SIM01}. These values give a 
crude estimate also for the $ 0\nu cc$ processes, although 
some  calculations indicate a larger coupling, \cite{STO03}.

At small $Q$ the capture amplitude  is given by the resonant 
conditions.  
Several  candidates for such a resonant capture may be found. 
Examples are given in Table II.
\begin{table}[htb]
\caption{The radiative neutrino-less capture rates from the 
$1S,2S$ states [1/y], $m_\nu = 1 $ eV. }
%\begin{center}
\begin{tabular}{ccccc}
 Atom & abundance $\%$  & $Q[keV] $ & $\Gamma(2S,1S)$     \\ \hline
  $^{92}_{42}$Mo  & 15.84 & 1628(5)      & $2 \cdot 10^{-32}$  \\
  $^{108}_{48}$Cd & 0.88  &  241(7)      & $1 \cdot10^{-31} $  \\
  $^{144}_{62}$Sm & 3.07  & 1730(4)      & $1\cdot 10^{-31} $  \\
  $^{162}_{68}$Er & 0.14  & 1781(4)      & $2\cdot 10^{-30} $  \\
  $^{180}_{74}$W  & 0.12  &  66(5)       & $2 \cdot10^{-31} $  \\
  $^{196}_{80}$Hg & 0.15  &  729(3)      & $4 \cdot 10^{-30}$  \\
\end{tabular}
%\end{center}
\end{table}
The region of practical interest is limited to $Q < 80$ keV. 
Such conditions are likely to be met in transitions  to 
excited states of final nuclei. The rate predictions are hampered 
by the poor knowledge 
of mass differences. Because of the sharp atomic resonances 
a few keV mass uncertainty results in rates changing by
several orders (cf. Table II). The atomic g.s. mass differences
$\Delta M$  
are taken from refs.\cite{TRE02},\cite{WAP03} and the $Q$ values 
are $\Delta M$  reduced by the  two electron-hole excitation energies.  
The precision needed is well below 1 keV.

The experimental signature for the double electron capture in 
the resonance conditions will be the $K$X - $K^h$X-ray coincidence 
and, in the case of decays to excited states, the triple 
coincidence with  the gamma rays 
deexciting these states.  The $K^h$ line has a measurable energy shift.
% The distinction between the $0\nu cc$ and $2\nu cc$ 
% transitions will in addition 
% rely on the calculated rates for the two processes: the very low 
% decay energy available for the neutrinos will strongly suppress the 
% $2\nu cc$ decay while enhancing the $0\nu cc$ one. 
The experimental feasibility arguments have to include the decay rate 
 and the cost estimates. Leaving the cost arguments aside and 
assuming 1 ton 
of the source material with 100 \%  isotopic purity and the 
correspondingly larger amount 
of a high resolution detector (e.g. a multilayered 
source-detector sandwich for X-ray detection combined with the 
medium resolution larger $\gamma$-ray detectors) it seems possible 
to design feasible experiments for the $0\nu$ double electron 
capture process. The count rates expected depend strongly on 
the energy difference with respect to the resonance value  
(see Table II).
The $^{112}$Sn isotope indicated in ref.\cite{BER83} as 
the best choice, now with the the recent mass determination  seems 
less profitable. Much higher decay rates can be expected for 
$^{152}$Gd and $^{164}$Er  $g.s.\rightarrow g.s.$ decays. 
However, there are no convenient experimental signatures in 
these transitions. Best signatures are offered in the cases of 
decays to excited states (e.g. $^{112}$Sn, 
$^{136}$Ce or $^{162}$Er) where there are characteristic 
high energy $\gamma$-rays in addition to the 
$k_{\alpha}$X-rays and the resonant transitions.

\begin{table}[htb]
\caption{The resonant situations. The radiative neutrino-less capture rates 
per year R/y, and per year and ton of the isotope  R/y $\cdot$ ton, 
 $m_{\nu} = 1 $ eV. 
The uncertainties of R/y $\cdot$ ton are due to  $1 \sigma$ errors in 
the mass determination. $Q$ - photon energies, $Q_{r}$ - the resonant 
energies,  $Q_{X}$,$Q_{\gamma}$ - energies of accompanying $X$-ray,
$\gamma$-ray photons,  $E^{*}$ - excitations of  final nuclei, [keV].}
% \cite{WAP04}}
% \begin{center}
\begin{tabular}{lllllll}
Atom & $^{112}_{50}$Sn & $^{136}_{58}$Ce & $^{152}_{64}$Gd & 
$^{162}_{68}$Er &  $^{164}_{68}$Er  & $^{180}_{74}$W \\  \hline
abnd.$\%$ & 1.01 & 0.19 & 0.20 & 0.14 & 1.56& 0.13  \\  
f.st. & $0^+_2$,1S,2P & $0^+_3$,1S,2P & g.s.,1S,2P & $1^+$,1S,2P& 
g.s.,2S,2P & g.s.,1S,2P \\
int.st. & $0^+_2$,1S,1S & $0^+_3$,1S,1S & g.s.,1S,2S & $1^+$,1S,1S& 
g.s.,2S,2S & g.s.,1S,1S \\
$\Delta M$ & 1919.5(4.6) & 2418.9(13) &54.6(3.5) & 1843.8(3.9)& 23.3(3.9)&
144.4(4.5) \\
$E^{*}$ & 1870.9 & 2315.4 &   & 1745.5  &        &      \\
$Q$ & 18.0(4.6) & 60(13) & 39.7(3.5) & 36.1(3.9) &6(4) & 68.6(4.5) \\
$Q_{r}$& 23.8      & 33     & 39.7      & 46.6      &1 & 56.3(4.5)  \\
$Q-Q_{r}$& -5.8(4.6)& 27(13)& 0.04(3.5) & -10.5(3.9)& 5(4)    & 12.3(4.5)  \\
$Q_{X}$ & 23   & 32     & 6.5     & 46  &  6.5    & 55  \\
$Q_{\gamma}$ & 1253.4   & 1496.9   &      &  1665.1   &   &   \\
$        $   & 617.6   &  818.5   &      &  80.6  &    &   \\
R/y & $7 \cdot 10^{-30}$ & $2 \cdot 10^{-30}$ & $2 \cdot 10^{-25}$ 
& $1 \cdot 10^{-28}$ & $2 \cdot 10^{-32}$ & $3 \cdot 10^{-28}$ \\
R/y $\cdot$ t & $1-10^{-2}$ & $10^{-1}-10^{-3}$ & $10^{3}-10^{-1}$ 
& $1-10^{-1}$ & $10^{-1}-10^{-3}$ & $3-.2$ \\
\end{tabular}
%\end{center}
\end{table}

\sc{The physical   background.}
 \normalshape
One advantage of the $0\nu \gamma $ process 
is a convenient  ratio of the signal to the physical 
background generated by the dominant $\nu \nu \gamma$ 
channel. Here, we briefly estimate this ratio defined as   
\begin{equation}
\label{b1}
R_{S/B} = \frac{\Gamma^{0 \nu \gamma}(Q) }{\Gamma^{\nu \nu \gamma}N_D}
\end{equation}  
where  $N_D$ is the fraction of photons from the dominant 
$\nu \nu \gamma $ decay mode emitted into the region  
from the end of the spectrum $Q$ down to $Q-D/2$ and $D$ is 
the photon energy resolution. 
More details may be found in ref. \cite{WYC03}. 
For easier comparison the two-neutrino radiative  rate is 
presented in terms of the no-neutrino radiative rate 
$\Gamma^{0 \nu \gamma }(k_{\gamma})$, given by eq.(\ref{a10}) or 
eq.(\ref{a11}),  as a function of the photon energy 
\begin{equation} 
\label{b2}
 \Gamma^{\nu \nu \gamma}= 
\int dL \delta(Q - \Sigma k_{i}) R_N^2
\Gamma^{0 \nu \gamma }(k_{\gamma}) \frac {(2\pi)^2}{k_{\gamma}}
\end{equation}  
where  the phase space element 
$ dL= \frac{2\pi d{\bf k_{\gamma}}d{\bf k_{\nu}}d{\bf k_{\nu}'}}
{(2\pi)^9 8 k_{\gamma} k_{\nu} k'_{\nu} }$ and 
$ R_N = (4\pi M^{2\nu})/(M^{0\nu} m_{\nu})$ is a dimensionless ratio 
of the nuclear two-neutrino and no-neutrino 
matrix elements. Typical values  $ M^{2\nu} \approx 1 $, 
$ M^{0\nu} \approx 1/fm $ follow from nuclear model calculations, 
\cite{EJI00},\cite{SUH98}.  
To obtain  $R_{S/B}$, expression (\ref{b2}) is presented as an integral 
over the photon energy distribution 
$ \Gamma^{\nu \nu \gamma}= \int_{0}^{Q} W( k_{\gamma} ) dk_{\gamma}$
and the background contribution becomes   
\begin{equation}
\label{b4}
\Gamma^{\nu \nu \gamma }N_D = \int_{Q-D/2}^{Q} 
W( k_{\gamma} ) dk_{\gamma}. 
\end{equation}
Formula $ W( k_{\gamma})= (Q-k_{\gamma})^3 \Gamma^{0 \nu \gamma }(k_{\gamma})
R_N^2/6(2\pi)^4 $  follows from eq.(\ref{b2}). This  indicates a  cubic 
 cut-off  at the end of the spectrum  due to the 
phase space. The  ratio $R_{S/B} $ is now given by the interplay of 
this cut-off and the  energy dependence of  
$\Gamma^{0 \nu \gamma }( k_{\gamma})$. 

First, we consider the magnetic type transition  related to 
the $1S,2S$ electron capture. 
For characteristic values  $ Q= 1$ MeV and $  m_{\nu} =1 $ eV one
obtains $ \Gamma^{0 \nu \gamma }/\Gamma^{\nu \nu \gamma }=  
5 \cdot 10^{-5}$. The photon energy resolution  $ D= 3$ KeV  
would yield very convenient $ R_{S/B} = 2 \cdot 10^{6}$. 
The case of resonant electron captures   is more involved 
due to the rapid energy dependence in $\Gamma^{0 \nu \gamma }(k_{\gamma})$   
The result is plotted 
in Fig.2. The  $  Q \leq  Q_{res}$  situation 
is very favorable and $ R_{S/B}$ is about $ 10^{9}$. 
For  $ Q > Q_{res}+ \Gamma^r $    
the  $R_{S/B}$ falls down as  $[ Q- Q_{res}]^{-5}$.   
At  $ Q- Q_{res} = 1 $ keV the   ratio is  
still very convenient, $ R_{S/B} \approx  10^{4}$,  but 
the conditions  deteriorate quickly with the increasing 
energy separation.

\begin{figure}
 \includegraphics[height=.3\textheight, width=.5\textwidth]{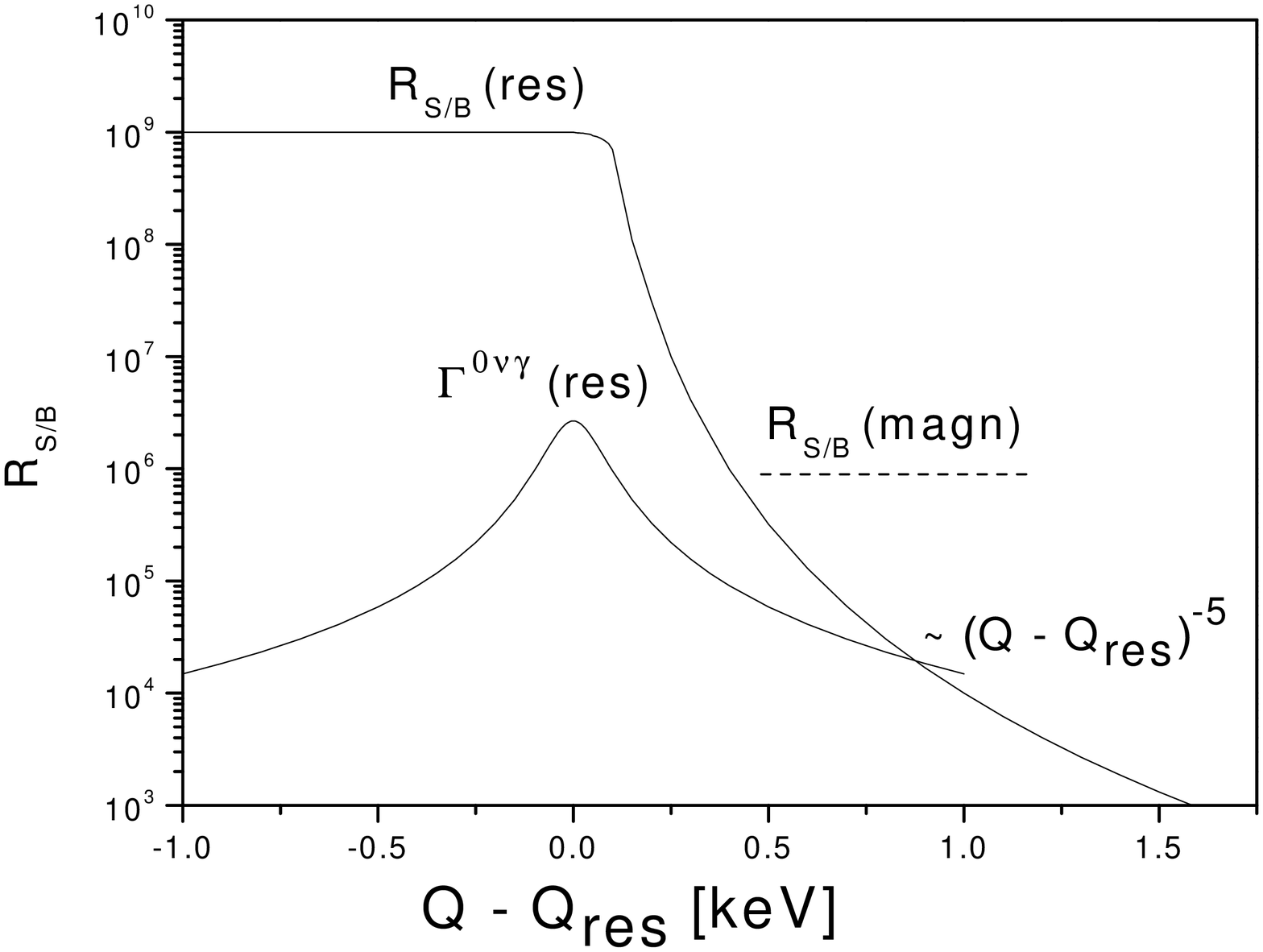}
 % \includegraphics{resonans02.eps}
 % \hfill
 % \includegraphics[height=.2\textheight, width=.2\textwidth]{neutrino2.eps}
%  \reflectbox{\includegraphics[height=.3\textheight,angle=-90]{neutron1.eps}}
% \includegraphics[angle=90, height=.05\textheight , width=.05\textwidth]{resonans02.eps}
 % \vspace{0.5cm}
 \caption{Schematic plot for the   signal/physical background ratios:  
 $ R_{S/B}(res)$- continuous line for the resonant transition and 
 $ R_{S/B}(magn)$- dashed line for the magnetic transition.
 The rate dependence $\Gamma^{0 \nu \gamma }(res)$  vs $(Q- Q_{res}) $ 
 is shown in arbitrary units to indicate the resonance width. } 
 \end{figure}

The random background  (RB)  varies  with the source material. Crudely, 
the effect/BR   ratios are expected to be improved in comparison 
to the calorimetric $\beta^-\beta^-$experiments \cite{KLA01} by a factor
between 10 and 1000, depending on the X-ray and the $\gamma$-ray 
coincidence trigger.

We conclude that detecting the neutrino-less double beta process
of any kind, be it $\beta ^-$, $\beta ^+$, or EC, will mean  discovering  the
lepton number nonconservation and proving the Majorana nature of
neutrinos. It will also provide a sensitive measure of the neutrino mass,
thus giving answers to some of the most urgent questions of to-day
physics. This search calls for a massive, diversified experimental as well
as theoretical effort. The radiative neutrino-less double electron capture, 
proposed in this work, is a valid alternative to double $\beta ^{\pm}$  emission, with several
experimental advantages. The predicted resonance rate enhancement makes
this search plausible. Exact rate predictions require precise
measurements of masses and likeways calculations of  nuclear
matrix elements. The former are feasible with 
modern techniques \cite{FRI02}, \cite{WAP03}. The latter are strongly assumption
dependent. The consensus \cite{NAR03} is that they can be trusted to about a
factor 3. One way to minimize this uncertainty is to carry out
measurements for several nuclear species, preferably on both sides of the
mass parabolae.  Thus the double electron capture search can be considered
not only as an attractive alternative to the double $\beta ^-$. Assuming both
kinds of experiment successful, it will also provide a much needed complementary 
piece of information.

\end{document}